**17/11/16**

# Overview of recent physics results from MAST

A Kirk[1], J Adamek[2], RJ Akers[1], S Allan[1], L Appel[1], F Arese Lucini[3], M Barnes[4], T Barrett[1], N Ben Ayed[1], W Boeglin[5], J Bradley[6], P K Browning[3], J Brunner[7], P Cahyna[2], S Cardnell[3], M Carr[1], F Casson[1], M Cecconello[8], C Challis[1], IT Chapman[1], S Chapman[13], J Chorley[7], S Conroy[8], N Conway[1], WA Cooper[9], M Cox[1], N Crocker[10], B Crowley[1], G Cunningham[1], A Danilov[11], D Darrow[12], R Dendy[13], D Dickinson[14], W Dorland[15], B Dudson[14], D Dunai[16], L Easy[14], S Elmore[1], M Evans[3], T Farley[6], N Fedorczak[17], A Field[1], G Fishpool[1], I Fitzgerald[1], M Fox[4], S Freethy[1], L Garzotti[1], YC Ghim[18], K Gi[19], K Gibson[14], M Gorelenkova[12], W Gracias[20], C Gurl[1], W Guttenfelder[12], C Ham[1], J Harrison[1], D Harting[1], E Havlickova[1], N Hawkes[1], T Hender[1], S Henderson[1], E Highcock[4], J Hillesheim[1], B Hnat[13], J Horacek[2], J Howard[21], D Howell[1], B Huang[1], K Imada[14], M Inomoto[19], R Imazawa[22], O Jones[7], K Kadowaki[19], S Kaye[12], D Keeling[1], I Klimek[8], M Kocan[23], L Kogan[1], M Komm[2], W Lai[13], J Leddy[14], H Leggate[24], J Hollocombe[1], B Lipschultz[14], S Lisgo[23], YQ Liu[1], B Lloyd[1], B Lomanowski[7], V Lukin[25], I Lupelli[1], G Maddison[1], J Madsen[26], J Mailloux[1], R Martin[1], G McArdle[1], K McClements[1], B McMillan[13], A Meakins[1], H Meyer[1], C Michael[21], F Militello[1], J Milnes[1], AW Morris[1], G Motojima[27], D Muir[1], G Naylor[1], A Nielsen[28], M O'Brien[1], T O'Gorman[1], M O'Mullane[29], J Olsen[26], J Omotani[1], Y Ono[30], S Pamela[1], L Pangione[1], F Parra[4], A Patel[1], W Peebles[10], R Perez[5], S Pinches[23], L Piron[1], M Price[1], M Reinke[14], P Ricci[9], F Riva[9], C Roach[1], M Romanelli[1], D Ryan[1], S Saarelma[1], A Saveliev[31], R Scannell[1], A Schekochihin[4], S Sharapov[1], R Sharples[7], V Shevchenko[1], K Shinohara[22], S Silburn[7], J Simpson[1], A Stanier[3], J Storrs[1], H Summers[29], Y Takase[30], P Tamain[17], H Tanabe[19], H Tanaka[32], K Tani[33], D Taylor[1], D Thomas[14], N Thomas-Davies[1], A Thornton[1], M Turnyanskiy[1], M Valovic[1], R Vann[14], F Van Wyk[4], N Walkden[14], T Watanabe[19], H Wilson[14], M Wischmeier[34], T Yamada[35], J Young[36], S Zoletnik[16] and the MAST Team and the EUROfusion MST1 Team[37]

1 CCFE, Culham Science Centre, Abingdon, Oxon, OX14 3DB, UK
2 Institute of Plasma Physics AS CR vvi, Prague, Czech Republic
3 Jodrell Bank Centre for Astrophysics, University of Manchester, Manchester M13 9PL, UK
4 Rudolf Peierls Centre for Theoretical Physics, University of Oxford, Oxford, UK
5 Department of Physics, Florida International University, Miami, FL 33199 Florida, USA
6 Department of Electrical Engineering and Electronics, University of Liverpool, Liverpool, UK
7 Department of physics, University of Durham, Durham, DH1 3LE, UK
8 VR, Uppsala University, SE-75120 Uppsala, Sweden
9 CRPP, EPFL, 1015 Lausanne, Switzerland
10 University of California, Los Angeles, Los Angeles, California 90095, USA
11 Russian Research Centre, Kurchatov Institute, Institute of Nuclear Fusion, Moscow, Russia
12 Princeton Plasma Physics Laboratory, PO Box 451, Princeton, New Jersey 08543, USA
13 Centre for Fusion, Space and Astrophysics, Department of Physics, Warwick University, UK
14 York Plasma Institute, Department of physics, University of York, Heslington, York, UK
15 University of Maryland, College Park, MD, USA
16 KFKI-RMKI, Pf. 49, H-1525 Budapest, Hungary

17 CEA, IRFM, F-13108 Saint Paul-lez-Durance, France
18 National Fusion Research Institute, Daejeon 169-148, Korea
19 Graduate School of Frontier Sciences, University of Tokyo, Tokyo, 113-0032, Japan
20 Departamento de Física, Universidad Carlos III de Madrid, 28911 Leganes, Spain
21 Plasma Research Laboratory, Australian National University, Canberra, ACT 0200, Australia
22 Japan Atomic Energy Agency, Ibaraki, 311-0193, Japan
23 ITER Organization, CS 90046, 13067 St Paul-Lez-Durance Cedex, France
24 Dublin City University, Glasnevin, Dublin, Ireland
25 National Science Foundation, Arlington, VA 22230, USA
26 Department of Physics, Technical University of Denmark, 2800 Kgs. Lyngby, Denmark
27 NIFS, Oroshi-cho, Toki City, Gifu, Japan
28 Risø, National Laboratory for Sustainable Energy, PO Box 49, Roskilde, Denmark
29 Department of Physics SUPA, University of Strathclyde, Glasgow, G4 ONG, UK
30 University of Tokyo, Kashiwa 277-8561, Japan
31 Ioffe Institute, Politekhnicheskaya 26, 194021 St. Petersburg, Russia
32 Graduate School of Energy Science, Kyoto University, Kyoto 606-8502, Japan
33 Tokyo Institute of Technology, Ookayama Campus, 2-12-1 Ookayama, Meguro-ku, Tokyo 152-8550, Japan
34 Max-Planck Institut fur Plasmaphysik, Boltzmannstrasse 2, D-85748 Garching, Germany
35 Faculty of Arts and Science, Kyusyu University, Fukuoka, 819-0395, Japan
36 University of Manchester, Manchester, UK
37 See appendix of H. Meyer et.al. (OV/P-12) Proc. 26th IAEA Fusion Energy Conf. 2016, Kyoto, Japan

## Abstract

New results from MAST are presented that focus on validating models in order to extrapolate to future devices. Measurements during start-up experiments have shown how the bulk ion temperature rise scales with the square of the reconnecting field. During the current ramp up models are not able to correctly predict the current diffusion. Experiments have been performed looking at edge and core turbulence. At the edge detailed studies have revealed how filament characteristic are responsible for determining the near and far SOL density profiles. In the core the intrinsic rotation and electron scale turbulence have been measured. The role that the fast ion gradient has on redistributing fast ions through fishbone modes has led to a redesign of the neutral beam injector on MAST Upgrade. In H-mode the turbulence at the pedestal top has been shown to be consistent with being due to electron temperature gradient modes. A reconnection process appears to occur during ELMs and the number of filaments released determines the power profile at the divertor. Resonant magnetic perturbations can mitigate ELMs provided the edge peeling response is maximised and the core kink response minimised. The mitigation of intrinsic error fields with toroidal mode number $n>1$ has been shown to be important for plasma performance.

## 1. Introduction

The Mega Ampere Spherical Tokamak (MAST) is a low aspect ratio device (R/a = 0.85m/0.65m ~ 1.3) with a cross-section similar to other medium sized devices. MAST's high-resolution diagnostic capability is complemented by sophisticated numerical modelling to facilitate deeper understanding. Although MAST has not operated since 2013 there has been substantial analysis and modelling performed on data obtained previously. The main aim of the analysis has been to validate models in order to allow extrapolation to future devices, in particular, MAST Upgrade [1], which is currently in the final stages of construction and will begin operation in 2017. Particular attention will be given to the areas of scenario development, fast particle physics and plasma exhaust, for which MAST Upgrade has unique capabilities. The layout of the paper follows the natural shot cycle. In section 2 start-up studies are described, while in section 3 we discuss how the current ramp can be used to test models of current diffusion. Section 4 is concerned with L-mode physics and in particular the Scrape Off Layer (SOL), core and fast ion physics. In section 5 we consider H-mode, specifically the nature of the pedestal, how the toroidal mode number of Edge Localised Modes (ELMs) affect the area over which the power is deposited on the divertor and how ELMs can be controlled using Resonant Magnetic Perturbations (RMPs). Section 6 looks at understanding and correcting intrinsic error fields which limit both the performance and duration of shots, while section 7 presents a summary and a look ahead to MAST Upgrade.

## 2. Start-up

MAST often used a so called "Merging-Compression" plasma start-up scheme in which the flux generated by ramping the current through an in-vessel low field side poloidal field coil (P3) is used to initiate the plasma. Breakdown occurs in the form of plasma rings around the P3 coils at the top and bottom of the vessel. The two current-carrying plasma rings approach each other due to mutual attraction, forming a current sheet and subsequently merge through magnetic reconnection into a single plasma torus, with substantial plasma heating. Detailed 2D profile measurements of electron and ion temperature and electron density have been made during merging reconnection start-up. The electron temperature forms a highly localized hot spot at the X-point, whilst the ion temperature increases downstream of this point [2]. When the toroidal field is more than three times the reconnecting field, the closed flux surfaces formed by the reconnected field sustain the temperature profile for longer than the electron/ion energy relaxation time ~4-10ms; with both profiles forming a triple peak structure centred on the X point. An increase in the toroidal field results in a more peaked electron temperature profile at the X-point (i.e. localised electron heating), and also produces higher ion temperatures at this point, but the ion temperature profile in the downstream region is unaffected [2]. The bulk ion temperature rise resulting from the reconnection process scales with the square of the reconnection field [3]. The ions are mostly heated in this downstream region by viscous dissipation and shock-like compressional damping of the outflow jet. A model based on magnetic helicity-conserving relaxation to a minimum energy state has been applied to the magnetic reconnection processes in both MAST and the solar corona [4]. In the case of MAST the average temperature rise predicted by this model is in good agreement with experimental measurements if it is assumed that most of the dissipated magnetic energy is converted to thermal energy. In addition, two-fluid (Hall Magneto Hydrodynamic) simulations of the merging process in MAST are able to match fairly well the measured temporal evolution of the density and temperature profiles.

## 3. Current ramp and current diffusion

Tokamaks typically use the current ramp-up to tailor the q-profile for the main heating phase of advanced tokamak scenarios. The capability to predict the q-profile evolution throughout the current ramp-up phase in response to the $I_P$ ramp-rate, externally applied heating, plasma density and plasma shaping is crucial for the design of new plasma scenarios. The q-profile evolution has been measured throughout the current ramp-up and flat-top phase of a plasma with no additional heating [5]. The current profile was determined by EFIT [6] equilibrium calculations constrained using high quality Motional Stark Effect (MSE) [7] measurements of the plasma, combined with high resolution Thompson scattering (TS) [8] and Z effective [9] measurements. The experimental data is used as inputs to the TRANSP [10] code, which models the current diffusion assuming neoclassical resistivity [11][12]. In both ramp-up and ramp-down experiments the current diffusion is not well modelled by TRANSP.

This inconsistency between simulation and experiment can be seen in Figure 1a-c. Figure 1b and c show the resulting value of the safety factor (q) produced from an MSE constrained EFIT obtained at specific times in the discharge compared to the results obtained from the TRANSP simulation at, respectively, the magnetic axis ($q_0$) and the half radius ($q_{0.5}$). Despite a good match to measurements at run initialisation, after 50ms the simulated MSE data lies well outside the region bounded by the experimental error bars. The discrepancy between measurements and simulation is particularly pronounced at the half radius during the current ramp and early flat-top. The discrepancy becomes significant at the axis in the latter stages of the $I_P$ flat-top where the modelled current diffusion is faster than that observed experimentally. There is no observable Magneto Hydrodynamic (MHD) activity in these plasmas in the period shown.

Experiments repeated in the flat top phase (see Figure 1d-f) show that after ~ 200ms of the flat top period the TRANSP simulations, which include neoclassical resistivity and a sawtooth model, are able to well describe the current evolution. In summary it would appear that our fundamental understanding of current diffusion appears to be correct but at present the models do not accurately reproduce the current ramp up phase.

In the present work the q profile has been used to conveniently illustrate the discrepancy between experiment and modelling. The advantage of the MSE diagnostic is that it gives a time-dependent measurement across the plasma profile and by specifying an equivalent synthetic diagnostic in the modelling; a direct comparison with the measurements is possible without any additional processing or interpretation. Other quantities, such as the loop voltage profile can, and have been calculated for theses shots. In the case of loop voltage, whilst the profile may be calculated, only the surface value is comparable to an experimental quantity.

All such information is being studied and shows the discrepancy between the modelling and experiment in a consistent manner and depending on how the model is constrained the discrepancy can be shown in different quantities. For example, if the modelling is constrained such that the current profile evolves according to the poloidal field diffusion equation and a specified resistivity model, the discrepancy can be shown as a difference between measured and calculated MSE angles and a difference between measured and calculated loop voltage. If the current profile is constrained by MSE measurements and loop voltage also constrained by experimental measurements, the discrepancy is shown as a difference between calculated and inferred resistivity profile. It is therefore still unclear which part of the calculation leads to the discrepancy.

## 4. L-mode

### 4.1 SOL transport

Understanding filamentary transport across the scrape off layer is a key issue for the design and operation of future devices as it is crucial in determining the power loadings to the divertor and first wall of the machine. The recent MAST exhaust programme has focussed on investigating the basic mechanisms responsible for setting plasma profiles in the SOL and their interplay with intermittent fluctuations (filaments), and on the physics of advanced divertors. A detailed characterisation of the MAST Scrape Off Layer has been performed including results from new diagnostics giving plasma potential and ion temperature measurements, which have then been compared to extensive modelling using the BOUT++ [13] and SOLPS [14] codes.

In L- and H-mode, it is commonly observed that the midplane density decay length increases with distance from the separatrix and with line averaged density. On MAST this density broadening occurs in the absence of detachment and independently from ionisation sources in the SOL [15]. In addition at similar densities, discharges with a higher plasma current do not show broadening, possibly due to the reduction in the connection length; implying that parallel as well as cross field transport regulate the SOL decay lengths [15]. Precise measurements of the density and electron temperature were also made available by a new binning technique of the Thomson scattering data. Mean profiles showed near-SOL decay lengths decrease with plasma current and increase with fuelling levels. A good correlation was found between decay lengths and the $D_\alpha$ emission, suggesting a role for the neutral particles in setting the profiles.

At the midplane, new measurements techniques, including a Ball-pen (BPP) [16] and a Retarding Field Energy Analyser (RFEA) [17] have been used to make profile measurements by attaching them to a reciprocating probe drive. The BPP technique was used to make profile measurements of plasma potential, electron temperature and radial electric field in L-mode plasmas [16]. The measured plasma potential profile is shown to significantly differ from the floating potential both in polarity and profile shape. By combining the BPP potential and the floating potential, the electron temperature was calculated and found to be in good agreement with the values obtained from the TS diagnostic when secondary electron emission is accounted for in the floating potential [16]. The calculated radial electric field ($E_R$) is of the order of 1kV/m and increases with plasma current. The simultaneous measurement of $E_R$ and the fluctuation characteristics from ion saturation current measurements ($I_{SAT}$) allows the birth location of the filaments to be investigated. Figure 2 shows a plot of the skewness of the $I_{SAT}$ signal versus the gradient of the electric field obtained during the reciprocation of the BPP into an L-mode plasma. The location where the skewness is smallest is often considered to be the region where the filaments are born and this is found to correspond to the location were the E-field gradient is largest.

Since plasma filaments play the dominant role in anomalous transport, they could also be important in determining the SOL profiles. At the target, L-mode filaments form spiral patterns that produce bands of increased heat flux, which were measured using infrared thermography [18]. These results showed that filaments can account for the full divertor target heat flux in the far SOL. In the near SOL the filaments overlap and hence it is more difficult to establish their contribution from the target data. To try to determine the role of filaments in the near SOL a study has been performed of the characteristics of the filaments as a function of plasma current observed in L-mode plasmas, using visible imaging at the mid-plane [19]. This analysis showed that the radial velocity ($V_r$), and to a lesser extent the radial size of the filaments, decreases as the plasma current is increased at constant density and input power. The decrease in $V_r$ is in the same proportion to the decrease in mid-plane density fall off length ($\lambda_{ne}$) namely; a 56% reduction in $V_r$ is consistent with a ~60% reduction in $\lambda_{ne}$ over the $I_P$ range studied (400-900 kA). Figure 3 demonstrates the observed correlation between $\lambda_{ne}$ and $V_r$, which is compatible with the idea that the filaments are responsible for determining the particle profiles at the divertor.

Detailed measurements of the motion, shape and amplitude of individual filaments identified in these high speed movies were compared to large scale 3D two-fluid simulations conducted in the STORM module of BOUT++ [20]. The simulations are able to reproduce the motion of the observed filaments within the accuracy of the experimental measurements. The numerical results showed that filaments characterised by similar size and light emission intensity can have quite different dynamics if the pressure perturbation is distributed differently between density and temperature components. In particular it has been found that a filament with a larger temperature perturbation moves slower in the radial direction than a filament dominated by a density perturbation [21]. This suggests a possible mechanism for the observed $I_P$ dependence discussed above whereby at higher $I_P$ the filaments have a higher $T_e$ and hence smaller $V_r$.

Linking the transient events to the midplane profiles, a theoretical framework was developed to interpret the experimental features of the density profiles on the basis of simple properties of the filaments, such as their radial motion and their draining towards the divertor [22]. This describes L-mode and inter-ELM filaments as a Poisson process in which each event is modelled as a wave function with (initial) amplitude and width statistically distributed according to experimental observations and dynamically evolving according to reduced fluid equations. The main strength of this statistical framework is its flexibility and its ability to test different models of filament dynamics. For example, it can be used to investigate the non-exponential nature of SOL profiles. It is found that a number of mechanisms may be able to explain the flattening of the density and electron temperature in the far SOL [23] and it is likely that several mechanisms are involved at the same time.

In order to guide experimental planning for MAST Upgrade the SOLPS code has been benchmarked against MAST discharges. The parameters extracted from this benchmarking have then been used to simulate both conventional and Super-X configurations in MAST-U [24]. In comparison to MAST, MAST-U will operate with a closed pumped divertor system (see Figure 4). MAST-U is designed to investigate a range of divertor topologies that have different strike point position, connection length and flux expansion. The simulations show that the Super-X configuration will detach at lower density (×1/3) or higher power (×4) with respect to the conventional divertor. The new divertor is predicted to significantly reduce the target power load through magnetic geometry and baffling as the tight closure of the divertor region leads to a strong increase in neutral density with associated power losses [24].

In order to simulate inter-ELM H-mode plasmas, the SOLPS radial transport diffusivities were tuned to give an upstream radial e-folding length for density of $\lambda_{ne}$=6.3 mm and for electron temperature of $\lambda_{Te}$=16.8 mm, in line with values observed for MAST in H-mode [25]. This tuning was done on the same MAST grid as used in reference [24]. The resulting simulated energy flux density at the target was fitted with an Eich function [26] to give a radial e-folding length of $\lambda_q$=8.0 mm (mapped to the outer mid-plane), in line with MAST infra-red measurements [27]. The same radial transport diffusivities were then used on conventional and Super-X MAST-U grids and density scans were performed in the core scope scenario (5 MW heating power, turbo-pumps only). It was assumed that up to 30% of the heating power would be lost via imperfect absorption, while a further 20% was assumed to be spent rebuilding the pedestal between ELMs. Thus the power into the simulation grids was 2.5 MW. Figure 5 shows the total deuterium ion flux to the outer target as a function of the outer mid-plane separatrix electron density, in conventional and Super-X geometries. In these simulations, in H-mode without impurity seeding, access to detachment (corresponding to full rollover of the target ion flux) at realistic upstream densities was only possible in the super-X configuration.

### 4.2 Core physics

Tokamak plasmas can rotate even when there is no external input of momentum, and this so-called “intrinsic rotation” can spontaneously change direction with relatively small changes in plasma conditions [28][29][30]. Understanding the mechanisms that drive this rotation is important for predictions of future reactors where neutral beams will typically produce little rotation. A Doppler backscattering (DBS) system has been used to make measurements of the intrinsic rotation in L-mode plasmas, including the first observation of intrinsic rotation reversals in a spherical tokamak [31]. Experimental results were compared to a 1D model, which captures the collisionality dependence of the radial transport of toroidal angular momentum due to finite drift orbit effects on turbulent fluctuations [32]. The model is able to accurately reproduce the change in sign of core toroidal rotation, using experimental density and temperature profiles from shots with rotation reversals as inputs but no other free parameters. However, there are some examples at low density and low plasma current that clearly disagree, which indicates there may be additional physics at play

The DBS system has also been used to investigate core turbulence. Ion scale turbulence ($k_\perp \rho_i$ <1.0) is strongly suppressed in spherical tokamaks, so in order to understand the dominant transport mechanisms it is important to diagnose higher wavenumbers. The high-k ($7 < k_\perp \rho_i < 11$) wavenumber spectrum of density fluctuations has been measured for the first time using the DBS system [33]. The DBS implementation used two-dimensional (2D) steering; this enabled high-k measurements with DBS, at $k_\perp >$ 20 $cm^{-1}$ ($k_\perp \rho_i > 10$) for launch frequencies < 75 GHz. A power law decrease in fluctuation signal was measured $|n(k_\perp)|^2 \propto k_\perp^{-\alpha}$ with $\alpha$= 4.7±0.2 for $7 < k_\perp \rho_i < 11$, which is similar to the value of $\alpha$ expected from turbulent cascade processes ($\alpha$=13/3).

While ion scale turbulence is strongly suppressed by flow shear, weak ion scale turbulence has been observed in a range of shots using beam emission spectroscopy around $\Psi_N$~0.7 [34]. Linear GS2 flux tube simulations showed that all the modes are linearly stable. However, non-linear simulations show beyond a certain threshold and given a large enough initial perturbation, subcritical turbulence can be sustained. These simulations suggest that the ion scale turbulence in MAST in the presence of flow shear is subcritical. In normal turbulence regimes the heat flux is proportional to the amplitude squared of the fluctuations, however, in the subcritical regime, near the threshold, the turbulent heat flux increases because coherent structures become more numerous (but not more intense) [35].

Helium [36], carbon and nitrogen [37] impurity transport coefficients have been determined from gas puff experiments carried out during repeat L-mode discharges. A weak screening of carbon and nitrogen is found in the plasma core, whereas the helium density profile is peaked. Both carbon and nitrogen experience a diffusivity of the order of 10 $m^2s^{-1}$ and a strong inward convective velocity of ~40 $ms^{-1}$ near the plasma edge, and a region of outward convective velocity at mid-radius. Neoclassical and quasi-linear gyrokinetic simulations have been performed with NEO [38] and GKW [39] respectively. Neoclassical transport alone is sufficient to explain the observed impurity transport of each species within $\rho<0.4$, but cannot explain the magnitudes of the transport coefficients or trend in Z in the region $0.4 <\rho< 0.8$. Whilst the equilibrium flow shear is sufficient to stabilise Ion Temperature Gradient modes (ITGs) in this region, Trapped Electron Modes (TEMs) are not stabilised, suggesting they may be the source of the anomalous transport [37].

**4.3 NBI heating and fast ion physics**

Understanding the behaviour of fast particles and their interaction with MHD instabilities is crucial for the operation and control of burning plasmas in future devices. In Spherical Tokamaks due the low toroidal field the neutral beam particles can be super-Alfvenic, which makes it possible to study fast ion-driven instabilities that resemble those likely to be driven unstable by $\alpha$-particles in ITER in plasmas characterized by a large fast particle pressure. Previous studies on MAST have indicated that the measured neutron rate is often lower than TRANSP predictions and this deficit is due to Fast Ion (FI) redistribution [40]. A series of experiments has been carried out to assess the effects that resonant and non-resonant MHD instabilities have on fast ions using a comprehensive set of fast ion diagnostics: a tangential and vertical Fast Ion $D_\alpha$ spectrometer (FIDA) system, a $^{235}U$ fission chamber, a neutron collimated flux monitor and a charged fusion product detector array. The aim was to integrate the observations from all diagnostics with a suite of codes to provide a consistent description of the fast ion behaviour. Good agreement is found for sawtooth induced fast ion losses between the experimental measurements and TRANSP simulations including the implementation of the Kadomstsev model [41]. Whilst ad hoc anomalous diffusion and loss models in TRANSP reproduce the global and local neutron emission during periods of fishbone excitation [42] they do not reproduce the spectrally and spatially resolved measurements from the FIDA system, showing that these models do not fully capture the effects of these modes on the fast-ion distribution [43].

Fishbones are known to be driven by gradients in the fast ion distribution, and therefore it is possible to control the instability by optimising the fast ion pressure profile to suppress these modes and prevent the redistribution or loss of the fast ions themselves. Experiments on MAST have demonstrated the efficacy of this approach by vertically displacing the plasma to achieve off-axis NBI fast ion injection [40] or by changing plasma density or NBI power to vary the magnitude of the fast ion pressure [44]. A series of shots were performed using two on axis neutral beam sources at fixed input power with increasing plasma density [44]. As the density increased the fishbone activity was observed to decrease and the measured neutron rate divided by the neutron rate calculated by TRANSP tended to unity (see Figure 6a). The TRANSP calculations show that as the density increases the maximum gradient in fast ion distribution decreases (see Figure 6b). Since operating at high density is not always desirable an alternative way of reducing the fast ion gradient is to change the NBI geometry. On MAST this was achieved by shifting the plasma vertically such that both NBI source were now off axis. This reduced the fast ion gradient (Figure 6b), which effectively eliminated the fishbone activity and gave improved agreement with the TRANSP calculations (Figure 6a).

These results have led to design options for MAST Upgrade to allow access to a wide range of plasma parameters without significant fast ion redistribution. In the operation of MAST-U in 2017, the NBI system will consist of two injectors, one located with the same injection geometry as those previously installed on MAST (referred to hereafter as the “on-axis” position), the other located with a horizontal line-of-sight 650mm above the midplane (referred to hereafter as the “off-axis” position). This combination will provide efficient core heating, via the on-axis injector, and significant off-axis heating and NBI Current Drive (NBCD) via the off-axis injector. The off-axis NBCD is intended to allow a certain degree of control of the q-profile with the intention of creating plasma scenarios with current profiles tailored to avoid deleterious MHD modes and allow a detailed investigation into the relationship between beam deposition profiles, MHD mode activity and fast ion transport. The triangle in Figure 6b shows the fast ion gradient predicted from the TRANSP simulations, which suggests that this beam configurations should have a significant mitigating effect in terms of the fishbone mode drive and is therefore expected to allow effective plasma heating and current drive over a wider range of plasma conditions in MAST Upgrade.

## 5. H-mode

### 5.1 Pedestal

There is a strong correlation between pedestal and core performance and hence understanding which turbulent processes determine the parameters of the pedestal is crucial for extrapolation to future devices. The DBS system has been used to measure intermediate-k ($k_\perp \rho_I$ ~ 3 to 4) density fluctuations at the top of the pedestal during the inter-ELM period of H-mode discharges. A novel diagnostic technique combining DBS with cross-polarization scattering (CP-DBS) enabled magnetic field fluctuations to also be locally measured at similar wave numbers [45]. Both measurements, which are shown in Figure 7 as a function of time during the ELM cycle, were made in a series of high-β ($\beta_N$~4.0–4.5) plasmas with large type-I ELMs with a ~ 9 ms period. The density fluctuations go down in the first 0.5–1.0 ms after the ELM and then remain approximately constant between 1 and 2 ms, which corresponds to the period in which the density pedestal is rapidly increasing. The period between 2 and 3 ms sees a sharp rise in the fluctuations and this corresponds to the period when the rapid pedestal evolution is arrested. Since these fluctuations are at a higher wavenumber than kinetic ballooning modes (KBMs) (the largest KBM growth rates were at $k_\perp$ ~ 0.2 $cm^{-1}$) it suggests that the other modes are determining the pedestal evolution at this stage of the ELM cycle. In the remainder of the ELM cycle the density fluctuations reach a saturated level. In contrast to the different stages observed in the density fluctuations, the magnetic field fluctuations slowly increase after ~ 2ms, with perhaps saturation at the end of the ELM cycle.

Linear GS2 calculations show that both microtearing modes (MTM) and electron temperature gradient (ETG) modes are unstable at the pedestal top at similar wave numbers to the measurements (although with more overlap with ETG wave numbers). Similar to what has been found previously [46] KBMs are unstable lower in the pedestal (at larger wavelengths). Taking into account the diagnostic differences when operating the DBS system in the two modes, the inferred ratio of fluctuation levels from experiment was $(\delta B/B)/(\delta n/n)=1/20=0.05$ [45]. Table 1 shows a comparison of the experimental measurements with the GS2 simulations. Both the experimental wave number range and the fluctuation ratio are more similar to the linear characteristics of the ETG than the MTM. These results imply that intermediate-k fluctuations due to the ETG play a role in inter-ELM pedestal evolution on MAST.

The pedestal pressure measured has been shown to increase as the global plasma pressure increases [47]. It was observed that the electron pressure pedestal height increased by around 100% for around a 40% increase in plasma pressure. Stability analysis shows that the enhanced Shafranov shift at higher core pressure stabilizes the ballooning modes driven by the pedestal pressure gradient, consequently allowing the pedestal to reach higher pressures. In order to replicate the experimentally observed electron pedestal height, and the scaling of the temperature and density pedestal height and width, an edge ion dilution had to be included in the numerical pedestal stability analysis.

The Synthetic Aperture Microwave Imaging (SAMI) diagnostic, originally designed to passively study electron Bernstein emission [48] has been used to conduct proof-of-principle 2D DBS experiments of the edge plasma. SAMI has been used to actively probe the plasma edge using a wide (±40° vertical and horizontal) and tuneable (10–34.5 GHz) beam. Conventional DBS systems have to be aligned at a specific orientation so that their probing beams are perpendicular to the magnetic field at the scattering location. In the SAMI system a phased array of antenna allow the Doppler backscattering to be focused in any direction. The system has been benchmarked against the conventional DBS system and used to produce 2D Doppler measurements. The optimum backscattering occurs when the probing beam is perpendicular to the magnetic field line, so by measuring the location of the maximum blue and red shifted components of the signal, the pitch angle can be deduced as a function of probing frequency and hence location in the plasma [49]. The preliminary results show that, provided the scattered signal is strong enough, the pitch angle derived from this technique is in good agreement with the EFIT equilibrium [6].

### 5.2 ELM physics

The so called Type I ELM [50] is thought to result from the peeling-ballooning MHD instability [51]. While the particle and energy losses from the plasma due to a type I ELM have been measured on a range of devices, including detailed measurements of the changes to the plasma profiles over the ELM crash, there is no detailed quantitative understanding of how these losses occur. The SAMI system in passive mode has also revealed that ELMs are accompanied by intense bursts of microwave emission in the electron cyclotron (EC) frequency range. These bursts provide strong evidence for the transient presence in the edge plasma of highly supra-thermal electrons and imply acceleration of electrons parallel to the magnetic field [52]. Particle-in-cell simulations show that magnetic field-aligned energetic electron distributions, of the kind likely to result from parallel electric fields generated by ELMs, excite electrostatic waves in the electron cyclotron range. If converted to electromagnetic modes, these waves could account for the SAMI emission. Whilst soft X-ray and Thomson scattering data suggest that the fraction of accelerated electrons is small, these electrons may play a significant role in ELM dynamics and provide evidence that a reconnection process is occurring.

The ELM wetted area is a key factor in the peak power load during an ELM, as it sets the region over which the ELM energy is deposited. The deposited heat flux at the target is seen to have striations in the profiles that are generated by the arrival of filaments ejected from the confined plasma. The effect of the filaments arriving at the target on the ELM wetted area has been measured using infrared (IR) thermography at the divertor and compared with the toroidal mode number obtained from high speed visible imaging of the main plasma [53]. Type I ELMs are found to have toroidal mode numbers of between 5 and 15. An increasing number of filaments at the target produce an increase in the wetted area. Figure 8a shows that the ratio of the ELM wetted area to the inter-ELM wetted area increases with the ELM toroidal mode number. Modelling of the ELMs has been used to understand the variation observed and the effect of an increased mode number; the modelling replicates the trends seen in the experimental data and supports the observation of increased toroidal mode number generating larger target ELM wetted areas.

In the ITER baseline Q =10 (where Q is the fusion power gain factor = $P_{fusion}/P_{in}$) scenario, which has a plasma current of 15 MA, the expected natural ELM frequency is ~1Hz with each ELM having and ELM energy loss $\Delta W_{ELM}$ ~ 20 MJ [54]. In order to ensure an adequate lifetime of the divertor targets on ITER the maximum ELM energy flux that can be repetitively deposited is 0.5 $MJm^{-2}$ [55]. In order to achieve this some level of ELM mitigation will be required. One of the parameters that enters into the calculation of the level of mitigation required is the changes in the power deposition profile. Figure 8b shows a prediction for the required mitigated ELM frequency expressed as a fraction of the natural ELM frequency as a function of the increase in wetted area during the ELM compared to inter-ELM for the $I_P$=15 MA discharges on ITER [55]. As can be seen the level of mitigation required decreases with the increase in wetted area. Hence these new results showing that the wetted area depends on the mode number of the ELM suggest that if the ELMs in ITER have a lower mode number [56], because they are closer to the peeling boundary, then a higher level of ELM control may be required.

Simulations of the ELM crash performed using the non-linear MHD code JOREK [57] have shown how the non-linear coupling of different mode numbers could be a necessary ingredient for ELM dynamics. This mode coupling has been found to be essential in order to reproduce the experimentally observed structures [58]. Hence it may be possible that a lower mode number of the ELM instability can still produce a larger observed number of filaments and hence wetted area.

Ion temperature measurements have been made for the filaments arriving at the divertor during ELMs using a retarding field energy analyser (RFEA) in a fast sweep mode [59]. As well as showing that the ion temperature in the filaments is much larger than the electron temperature, they have revealed that in a certain category of ELMs the filaments arrive at the target over an extended time (>1 ms), compared to the normal duration of 200-300 μs. Similar to what has been seen previously on other devices (NSTX [60], JET [61], AUG [62]) these ELMs appear to be composed of primary and secondary filaments. Normally on MAST there is a large distance between the plasma and any structure. These secondary filaments only occur when the plasma shape is such that there is a large plasma interaction with an in vessel poloidal field coil (P3). This suggests that there may be a mechanism by which the primary filaments interact with in vessel components (or the first wall in other devices) that releases impurities or neutrals which then change the edge stability and lead to further filaments being released.

### 5.3 ELM control

All current estimations of the energy released by type I ELMs indicate that, in order to ensure an adequate lifetime of the divertor targets on ITER, a mechanism is required to decrease the amount of energy released by an ELM, or to eliminate ELMs altogether. One such amelioration mechanism relies on perturbing the magnetic field in the edge plasma region, either leading to more frequent, smaller ELMs (ELM mitigation) or ELM suppression (see [63] and references therein). On MAST it has been shown previously that although the ELM frequency ($f_{ELM}$) increases with the applied resonance field ($b^r_{res}$) above a certain threshold, this threshold depends on the toroidal mode number of the applied perturbation [64]. Calculations performed using the MARS-F code [65], which is a linear single fluid resistive MHD code that combines the plasma response with the vacuum perturbations, including screening effects due to toroidal rotation, show that the plasma response leads to plasma displacements normal to the flux surfaces [66][67]. The poloidal location of the maximum in the displacement can vary. The X-point displacement is largest when the edge peeling-tearing response dominants whereas the displacement at the midplane is largest when the core kink component dominates. Previous studies on MAST showed that ELM mitigation only occurs when the X-point displacement is larger than the mid-plane displacement [66] and similar results have also been obtained on ASDEX Upgrade [68]. A detailed scan performed in the last campaign on MAST has illustrated further the importance of minimising core-kink response. MAST had 6 coils in the upper row and 12 coils in the lower row. By adjusting the current in the lower ELM coils in an n=3 configuration it was possible to adjust the phase difference ($\Delta\phi$) between the field patterns in the upper and lower row of coils. Figure 9a shows a plot of $f_{ELM}$ versus $\Delta\phi$ for a series of repeat discharges in which $\Delta\phi$ was changed shot to shot. $f_{ELM}$ increases as $\Delta\phi$ is decreased from 0 to -90°. For discharges with $\Delta\phi>0$ the discharge is terminated by a locked mode soon after the RMP coil current reaches its maximum value. Figure 9b shows the resonant field component ($b^r_{res}$) in the vacuum approximation and taking into account the plasma response as a function of $\Delta\phi$. As observed on other devices [68] there is an offset in the $\Delta\phi$ location of the peak for the vacuum and plasma response. Based on this observation it could be concluded that the results were in better agreement with the vacuum than the plasma response. However, looking in more detail, Figure 9c reveals that there is only a small window in $\Delta\phi$ for which the X-point displacement is larger than the mid-plane displacement. It is only in this window that ELM mitigation is observed without producing a locked mode. Figure 9d shows that the peeling response is effectively flat for $-120<\Delta\phi<-30^{o}$. Although the peeling response increases for $\Delta\phi>0$ the kink response is also rising rapidly and dominates and this is presumably the cause of the locked modes observed in this region. Hence in order to achieve the best ELM mitigation it is necessary to maximise the peeling response and minimise the kink response.

The effect that the RMPs have on the confinement of energetic (neutral beam) ions has been investigated using measurements of neutrons, fusion protons and FIDA light emission [69]. In the worst case of a low plasma current (400 kA) discharge with RMPs applied with a toroidal mode number n = 3 the total neutron emission dropped by approximately a factor of two. Simulations of RMP-induced fast ion transport in MAST, using the F3D-OFMC code, have been able to reproduce these results [70]. For higher n RMPs and/or higher plasma current the losses were considerably lower.

## 6. Intrinsic error fields

The misalignment of field coils in tokamaks can lead to toroidal asymmetries in the magnetic field, which are known as intrinsic error fields. These error fields often lead to the formation of locked modes in the plasma, which limit the lowest density that is achievable. Measurements on MAST suggest that the dominant source of the intrinsic error field was due to the P4 and P5 poloidal field coils. Since these coils will be re-used in MAST-U a series of measurements were made to characterise the field structure they generate and a set of experiments performed to understand how best to minimise their effect. A direct measurement of the toroidal asymmetry of the fields from these coils has been made, which has then been parametrized in terms of distortions to the coils. Empirically, the error fields are corrected using error field correction coils, where the optimum correction is found by determining the current required to ensure that the discharge is furthest from the onset of a locked mode. Assuming that the dominant n = 1 error field is produced by the P4 and P5 coils, the empirically derived corrections have been compared with the known distortion of these coils [71]. In the vacuum approximation there is a factor of ~3 difference between the predicted and empirically determined correction. These studies have been extended to a comparison with full MHD plasma response calculations [72]. Various optimization criteria have been compared to the experimental results and the two which are most compatible with the data are one that aims to minimise the net toroidal resonant electromagnetic torque on the plasma column and the other corresponds to the full cancellation of the 2/1 resonant field harmonic at the q = 2 surface, including the plasma response. When the plasma response is included better agreement is obtained, but there are still some cases where the agreement is not good. The results suggest that other effects may be important. These include on the experimental side additional unmeasured sources of the error field or on the theory side the effect of other higher n toroidal harmonics or the non-linear coupling of the error field to the plasma.

Even if the n = 1 component can be corrected there is still a large residual n = 2 component from the P4 and P5 coils, which could be the source of the rotation braking in shot 29912 shown in Figure 10. This n=2 components can in principle be corrected using the RMP coils in an n=2 configuration. The optimum n=2 correction that minimised the **j**x**B** torque on the plasma was calculated using MARS-F and applied to a plasma that already had an optimised n=1 correction applied. Figure 10 shows that such a correction can reduce the plasma braking and improve the quality of the H-mode (determined by the regularity of the ELMs).

Rather than correcting the error fields the best policy is to reduce the error field to the lowest possible value during the build stage. For MAST-U the toroidal variation in the field from the coils has been measured and then the optimum installation angle and location of the coils has then been calculated so as to minimise the n=1 components of their intrinsic error fields.

## 7. Summary and future plans

New results from MAST are presented that focus on validating models in order to extrapolate to future devices. Detailed measurements during start-up experiments, combined with modelling, have shown how the bulk ion temperature increases as a result of the reconnection process and scales with the square of the reconnecting field. An analysis of the changes in the q profile as a function of time show that while our fundamental understanding of current diffusion appears to be correct, at present the models do not accurately reproduce the current ramp up phase. A detailed characterisation of the MAST SOL has been performed including results from new diagnostics giving plasma potential and ion temperature measurements. Detailed studies have revealed how filament characteristic are responsible for determining the near and far SOL density profiles. These measurements have been compared to extensive modelling, including 3D effects on filaments dynamics with the BOUT++ code, and benchmarking the SOLPS code.

A DBS system has been used to make measurements of both core and edge turbulence. Measurements of the intrinsic rotation in L-mode plasmas show good agreement with a model that captures the collisionality dependence of the radial transport of toroidal angular momentum due to finite drift orbit effects on turbulent fluctuations. Ion scale turbulence is strongly suppressed in MAST and the DBS system has been used to measure core turbulence at higher wavenumbers. The observed turbulence is consistent with being created by a turbulent cascade process. The density fluctuation measurements from the DBS have been combined with measurements of the magnetics fluctuations obtained with cross-polarisation DBS. When compared with GS2 simulations the experimental observations of the relative amplitudes and wavelengths of the density and magnetic field fluctuations at the top of the pedestal are more similar to the linear characteristics of electron temperature gradient modes than micro tearing modes.

Comprehensive measurements from a suite of diagnostics on MAST have shown the effect that core MHD modes and RMPs have on the confinement and redistribution of fast ions arising from neutral beam injection.  Subsequent experiments on MAST demonstrated that by vertically displacing the plasma to achieve off-axis NBI fast ion injection or by changing plasma density it is possible to vary the fast ion pressure gradient and mitigate the redistribution.

Microwave bursts have been observed at the onset of ELMs, which suggest that a reconnection process may be occurring, possibly resulting in the release of energy from the plasma. The number of filaments observed in an ELM is found to be correlated with the increase in extent of the power footprint at the divertor. This suggests that the level of ELM mitigation required on future devices, such as ITER, will depend on the toroidal mode number of the ELMs. Studies of ELM mitigation on MAST have shown the role that the plasma response plays in determining the optimum non-axisymmetric field configuration. The access criteria for ELM mitigation using RMPs has been found to require maximising the edge peeling response combined with minimising the core kink response. Whilst RMPs are essential for mitigating ELMs they do have consequences for the core confinement and in particular the confinement of fast ions, which are sensitive to the mode number of the applied perturbation.

Intrinsic error fields, produced due to toroidal asymmetries in the magnetic field coils, often lead to the formation of locked modes in the plasma limiting the lowest density that is achievable. It has been found that the plasma response to such 3D field perturbations must be included in modelling in order to get agreement with the observations. Typically only error field components that have a toroidal mode number n=1 are corrected, however, experiments have shown that the n=2 component can also lead to rotation braking and can and should be corrected.

The results presented here have been used to improve the design of MAST Upgrade, which will begin operation in late 2017 [1]. The MAST Upgrade research programme has three primary objectives: 1) To develop reactor-relevant advanced divertor concepts, 2) add to the knowledge base for ITER and 3) explore the feasibility of using a spherical tokamak as the basis for a fusion Component Test Facility. To deliver this capability the load assembly is being comprehensively upgraded in stages and the first stage, known as “core scope”, is now nearing completion. Core scope includes 17 new shaping and divertor poloidal field coils (14 inside the vessel), and a new closed pump-able divertor structure to make a highly flexible exhaust physics platform (see Figure 4). This stage of the upgrade will also provide a 50% increase in the toroidal field (from 0.585 to 0.92 T at R = 0.7m) and a near doubling of the inductive flux from the central solenoid (0.9 to 1.7Vs (1.6 Wb)), which should allow access to a plasma current of 2MA. One of the present neutral beams will be moved off-axis for improved current profile control and fast ion physics studies. It will be equipped with ELM control coils, many new diagnostics and an extensive gas fuelling system.

## Acknowledgement

We acknowledge the contributions of the EUROFusion MST1 team. This work has been carried out within the framework of the EUROfusion Consortium and has received funding from the Euratom research and training programme 2014-2018 under grant agreement No 633053 and from the RCUK Energy Programme [grant number EP/I501045]. To obtain further information on the data and models underlying this paper please contact PublicationsManager@ccfe.ac.uk. The views and opinions expressed herein do not necessarily reflect those of the European Commission.

## References

[1] Morris AW 2012 IEEE Transactions on Plasma Science **40** 682

[2] Tanabe H *et al.*, 2015 Phys. Rev. Lett. **115** 215004

[3] Tanabe H *et al.*, EX/P4-32 this conference

[4] Browning P *et al.*, 2016 Plasma Phys. Control. Fusion **58** 014041

[5] Keeling D *et al.*, Proc 35th EPS Conf. 2008; Turnyanskiy M *et al.*, 2009 Nucl. Fusion **49** 065002

[6] Lao LL et al., 1985 Nucl. Fusion **25** 1611

[7] Conway NJ *et al.*, 2010 Rev. Sci. Instrum. **81** 10D738

[8] Scannell R *et al.*, 2010 Rev. Sci. Instrum. **81** 10D520

[9] Patel A *et al*., 2004 Rev. Sci. Instrum. **75** 4944

[10] Goldston RJ et al., 1981 J. Comput. Phys. **43** 61

[11] Hinton FL and Hazeltine RD, 1976 Rev. Mod. Physics **48** 239

[12] Sauter O *et al.*, 1999 Phys. Plasmas **6** 2834; Sauter O *et al.*, 2002 Phys. Plasmas **9** 5140

[13] Dudson DB, Umansky M, Xu X, Snyder P and Wilson H, 2009 Comput. Phys. Commun. **180** 1467

[14] Schneider R et al 2006 Contrib. Plasma Phys. **46** 3

[15] Militello F et al., 2016 Nucl. Fusion **56** 016006

[16] Walkden NR et al., 2015 Rev. Sci. Instrum. **86** 023510

[17] Allan S et al., 2016 'Ion Temperature Measurements of L-mode Filaments in MAST by Retarding Field Energy Analyser' submitted to Plasma Phys. Control. Fusion

[18] Thornton AJ et al., 2015 Plasma Phys. Control. Fusion **57** 115010

[19] Kirk A et al., 2016 Plasma Phys. Control. Fusion **58** 085008

[20] Militello F et al., 2016 Plasma Phys. Control. Fusion **58** 105002

[21] Walkden NR et al., 2016 ‘Dynamics of 3D isolated thermal filaments’ Submitted to Plasma Phys. Control. Fusion

[22] Militello F and Omotani JT, 2016 Nucl. Fusion **56** 104004

[23] Militello F and Omotani JT, 2016 ‘On the relation between non-exponential Scrape Off Layer profiles and the dynamics of filaments’ Submitted to Plasma Phys. Control. Fusion

[24] Havlickova E et al, 2015 Plasma Phys. Control. Fusion **57** 115001

[25] Kirk A et al, 2004 Plasma Phys. Control. Fusion **46** 1591

[26] Eich T et al, 2011 Phys. Rev. Lett. **107** 215001

[27] Thornton A et al, 2014 Plasma Phys. Control. Fusion **56** 055008

[28] Bortolon A et al 2006 Phys. Rev. Lett. **97** 235003

[29] Rice JE 2011 Phys. Rev. Lett. **107** 265001

[30] Angioni C et al 2011 Phys. Rev. Lett. **107** 215003

[31] Hillesheim JC et al. 2015 Nucl. Fusion **55** 032003

[32] Barnes M et al., 2014 Phys. Rev. Lett. **111** 055005

[33] Hillesheim JC et al., 2015 Nucl. Fusion **55** 073024

[34] Field AR et al., 2014 Plasma Phys. Control. Fusion **56** 025012

[35] Van Wyk F et al., 2016 ‘Transition to subcritical turbulence in a tokamak plasma’ to be published in Journal of Plasma Physics

[36] Henderson S et al. 2014 Nucl. Fusion **54** 093013

[37] Henderson S et al. 2015 Plasma Phys. Control. Fusion **57** 095001

[38] Belli EA and Candy J, 2008 Plasma Phys. Control. Fusion **50** 095010.

[39] Peeters AG et al., 2009 Comput. Phys. Commun. **180** 2650

[40] Turnyanskiy M. et al 2013 Nucl Fusion **53** 053016

[41] Cecconello M et al. 2015 Plasma Phys. Control. Fusion **57** 014006

[42] Klimek I et al., 2015 Nucl. Fusion **55** 23003

[43] Jones OM et al. 2015 Plasma Phys. Control. Fusion **57** 125009

[44] Keeling DL *et al* 2015 Nucl. Fusion **55** 013021

[45] Hillesheim JC et al., 2015 Plasma Phys. Controlled Fusion **58** 014020

[46] Dickinson D et al., 2012 Phys. Rev. Lett. **108** 135002

[47] Chapman IT et al 2015 Nucl. Fusion **55** 0130041

[48] Freethy SJ et al., 2013 Plasma Phys. Controlled Fusion **55** 124010

[49] Thomas D *et al* 2016 Nucl. Fusion **56** 026013

[50] Zohm H 1996 Plasma Phys. Control. Fusion **38** 105

[51] Connor JW 1998 Plasma Phys. Control. Fusion **40** 531

[52] Freethy SJ et al. 2015 Phys. Rev. Lett. **114** 125004

[53] Thornton AJ et al 2016 'The role of ELM filaments in setting the ELM wetted area in MAST and the implications for future devices' Submitted to Plasma Phys. Controlled Fusion

[54] Loarte A et al 2003 Plasma Phys. Control. Fusion **45** 1549

[55] Loarte A et al 2014 Nucl. Fusion **54** 033007

[56] Snyder PB et al. 2011 Nucl. Fusion **51** 103016

[57] Huysmans G et al., 2007 Nucl. Fusion **47** 659

[58] Pamela S et al "Multi-machine modelling of ELMs and pedestal confinement:From validation to prediction" TH/8-2, paper presented at 26th IAEA Int. Conf. on Fusion Energy Kyoto, Japan 2016

[59] Elmore S et al 2016 Plasma Phys. Control. Fusion **55** 065002

[60] Maqueda R et al 2009 J. Nucl. Mater. **390–1** 843–6

[61] Silva C et al 2009 Plasma Phys. Control. Fusion **51** 105001

[62] Kirk A et al 2011 Plasma Phys. Control. Fusion **53** 035003

[63] Kirk A et al 2013 Plasma Phys. Control. Fusion **55** 124003

[64] Kirk A et al 2013 Plasma Phys. Control. Fusion **55** 115006

[65] Liu YQ et al 2010 Phys. Plasmas **17** 122502

[66] Liu YQ et al 2011 Nucl. Fusion **51** 083002

[67] Haskey S.R. et al 2014 Plasma Phys. Control. Fusion **56** 035005

[68] Kirk A et al 2015 Nucl. Fusion **55** 043011

[69] McClements KG et al 2015 Plasma Phys. Control. Fusion **57** 075003

[70] Tani K et al. 2016 ``Application of a non-steady-state orbit-following Monte- Carlo code to neutron modeling in the MAST spherical tokamak'', Plasma Phys. Control. Fusion, in press.

[71] Kirk A et al 2014 2015 Plasma Phys. Control. Fusion **56** 104003

[72] Liu YQ et al 2014 2015 Plasma Phys. Control. Fusion **56** 104002

## Tables

**Table 1** Comparison of wavenumber $\boldsymbol{k}\perp$ and the ratio of magnetic to density fluctuations $(\tilde{\boldsymbol{B}}/\boldsymbol{B})/(\tilde{\boldsymbol{n}}/\boldsymbol{n})$ measured at the top of the pedestal and from GS2 simulations for MicroTearing Modes (MTM) and Electron Temperature Gradient (ETG) modes.

| | $k\perp$ (cm$^{-1}$) | $(\tilde{B}/B)/(\tilde{n}/n)$ |
|---|---|---|
| Experiment | 6-9 | 0.05 |
| MTM | 0.5-4.0 | 0.4 |
| ETG | 4.0-30.0 | 0.02 |

## Figures

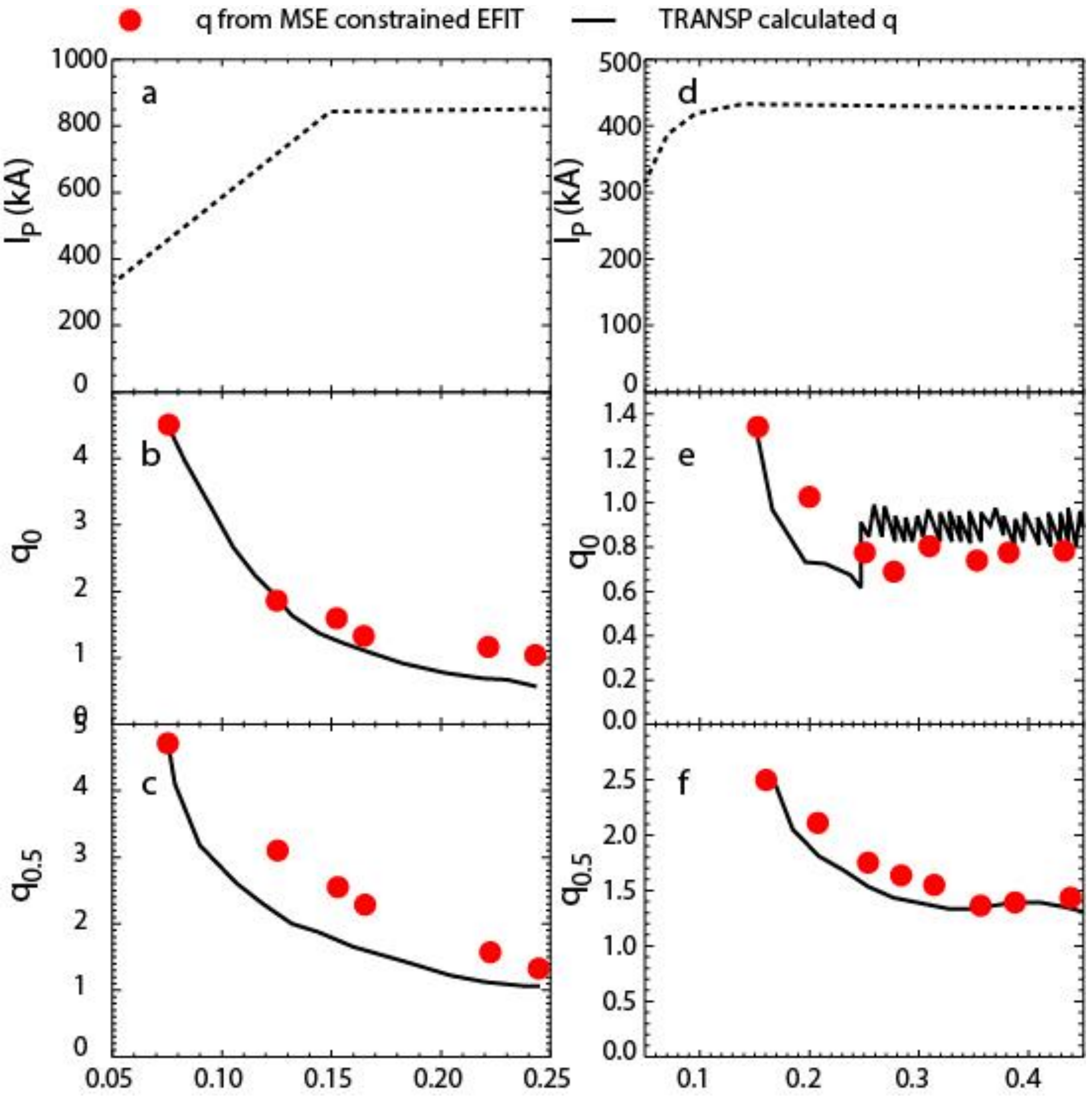

**Figure 1** q profile evolutions during $I_P$ ramp-up: a) $I_P$ waveform, b) and c) $q_0$ and $q_{0.5}$ traces from MSE constrained EFIT and TRANSP simulations. q profile evolutions during $I_P$ flat-top: d) $I_P$ waveform, b) and c) $q_0$ and $q_{0.5}$ traces from MSE constrained EFIT and TRANSP simulations.

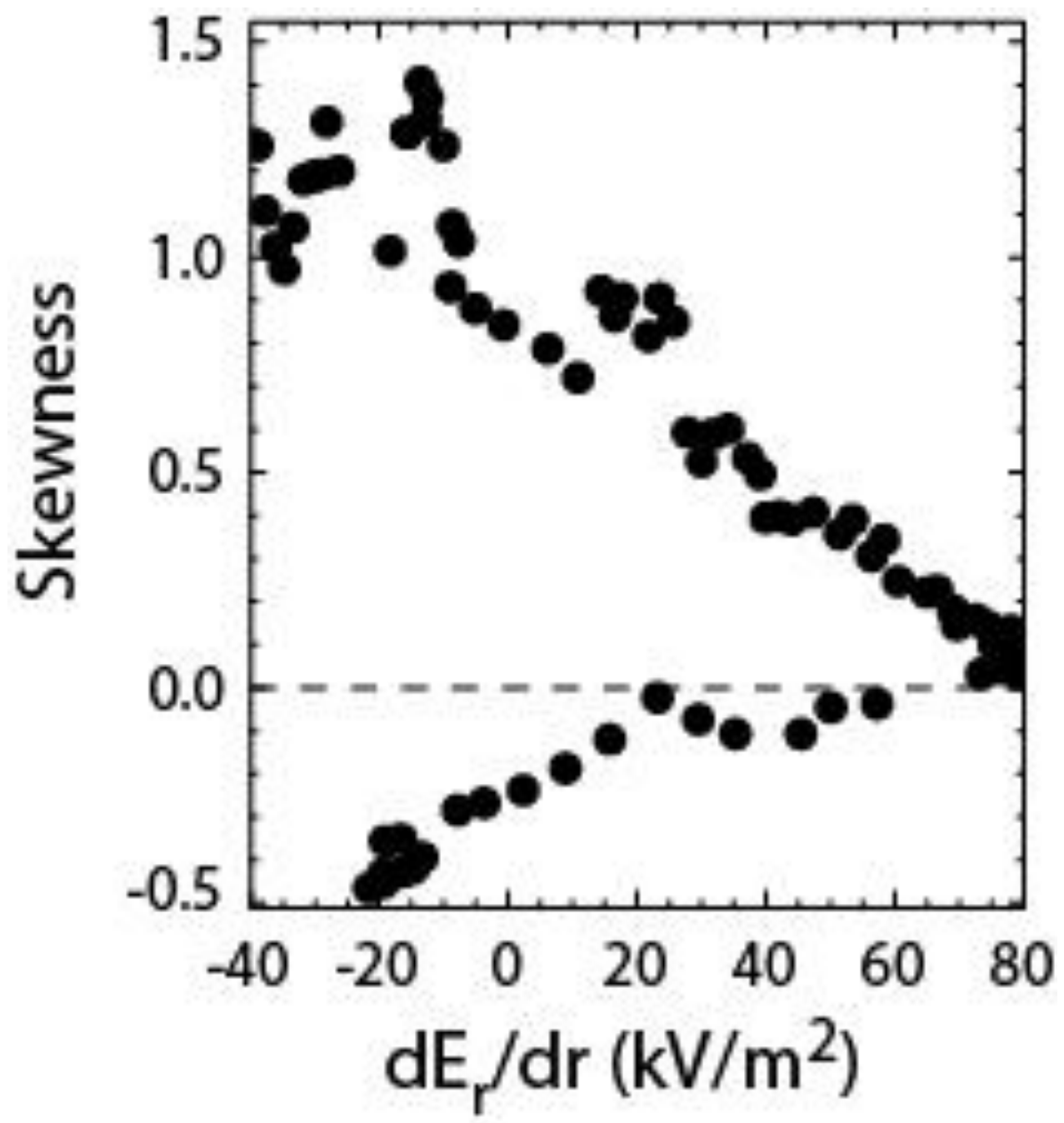

**Figure 2** The skewness of the ion saturation current versus the gradient in the radial electric field from ball pen probe measurements during a reciprocation into the edge of an L-mode plasma.

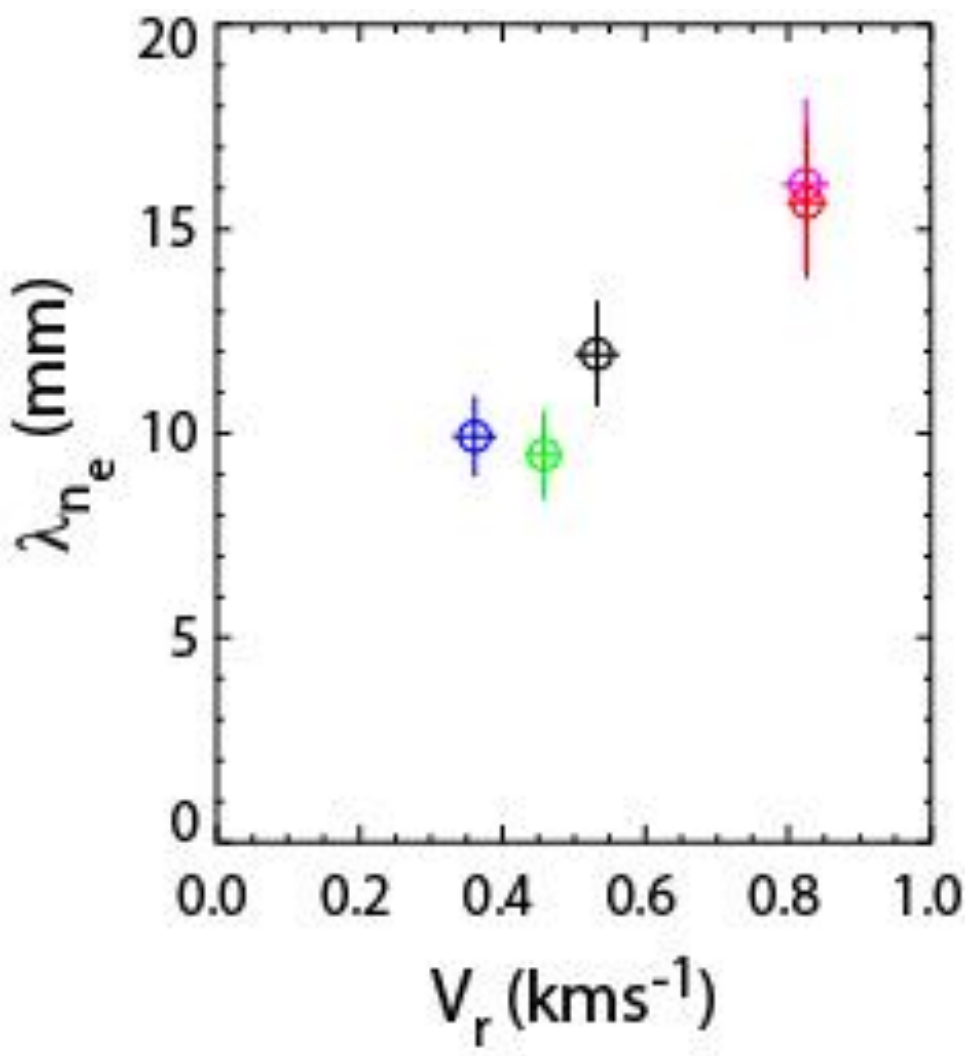

**Figure 3** The mid-plane density fall off length ($\lambda_{ne}$) as a function of the mean radial velocity of the filaments ($V_r$) determined from a range of shots at different plasma current.

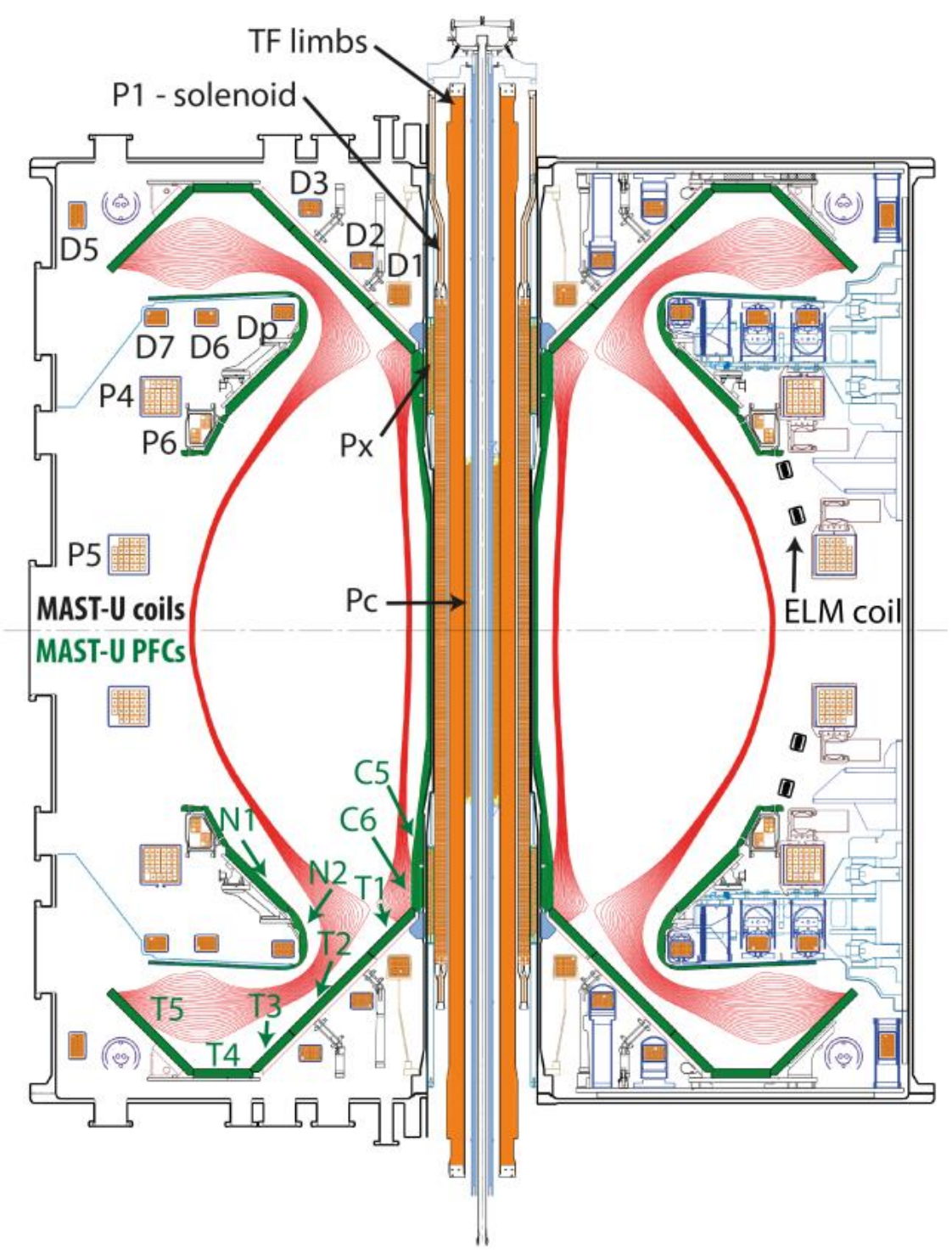

**Figure 4** Cross-section of MAST-U showing the location of the shaping coils and divertor tiles. Superimposed are magnetic field lines in a Super-X configuration.

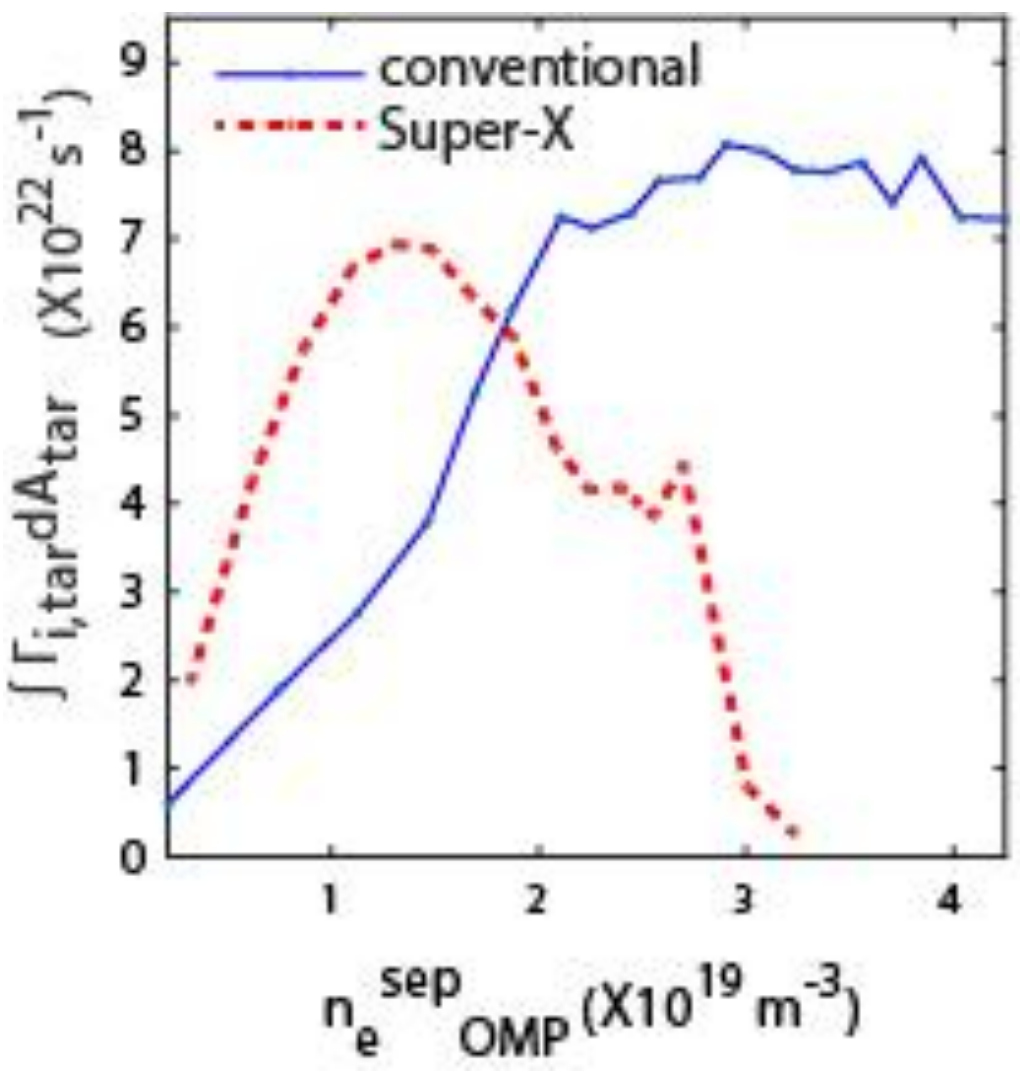

**Figure 5** SOLPS Simulation results of the total ion flux to the target versus separatrix density at the midplane for an inter ELM period of a H-mode plasma in MAST-U with the divertor in a conventional (solid) and Super-X configuration (dashed).

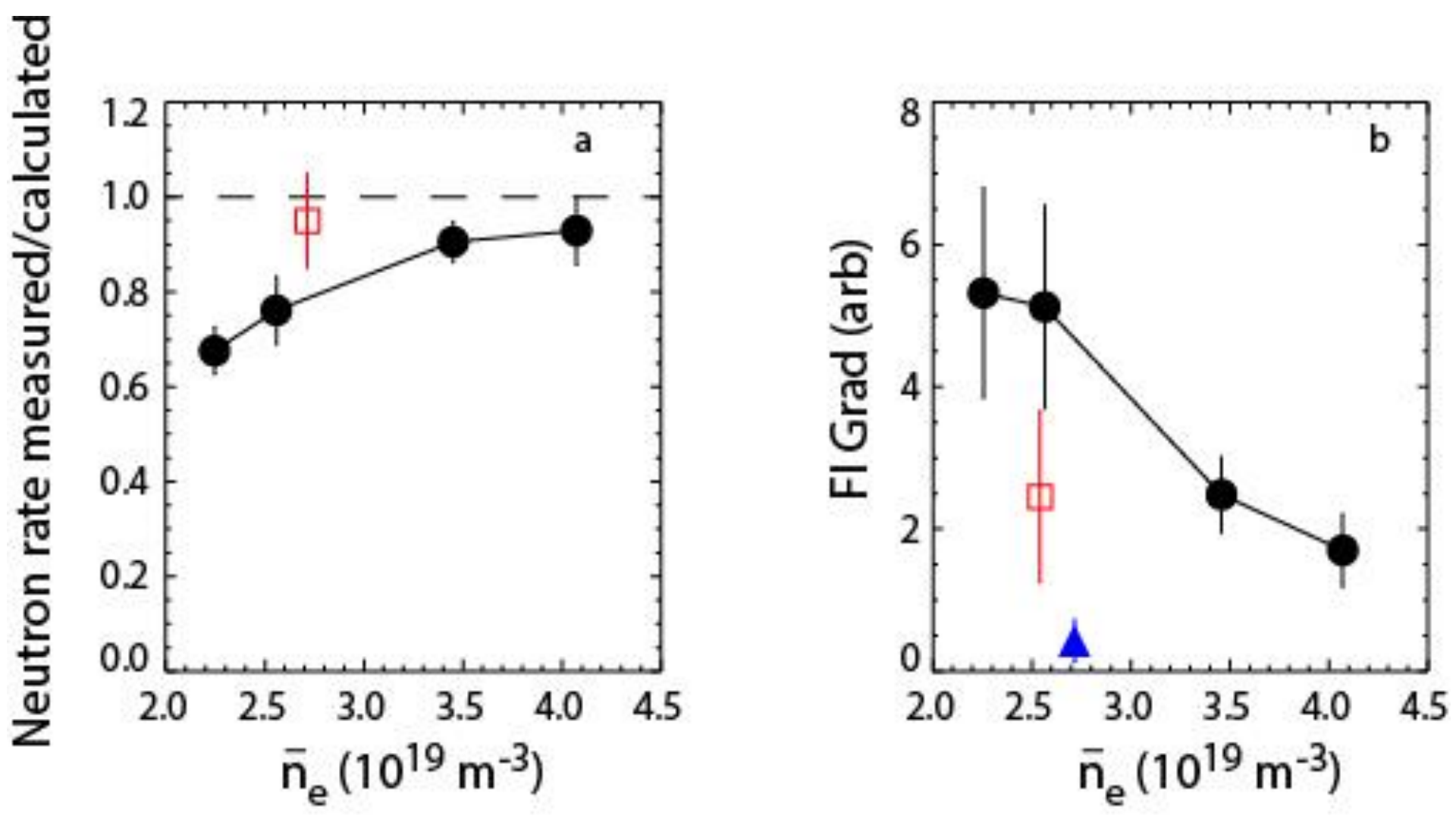

**Figure 6** a) Neutron emission rate measured by the Fission chamber (FC) divided by TRANSP calculated neutron emission rate and b) maximum fast ion gradient ($\partial f / \partial \psi_{p,N}$) (in arbitrary units) versus line-averaged density for shots with two beams on axis (circles) and vertically shifted (Square). The triangle in b) shows a calculation for a MAST-U shot with one on and one off-axis beam

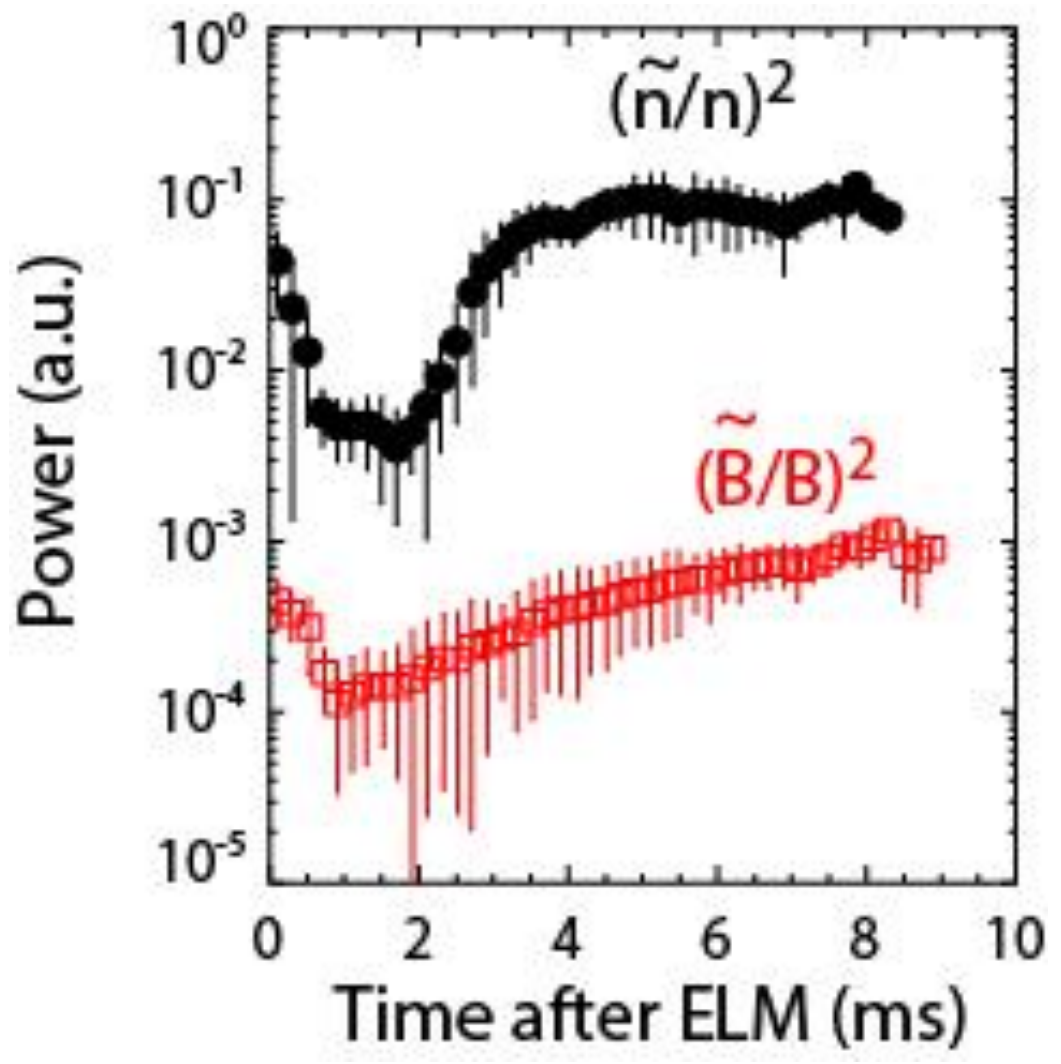

**Figure 7** DBS measurements of the density fluctuations $(\frac{\tilde{n}}{n})^2$ and CP-DBS measurements of the magnetic fluctuations $(\frac{\tilde{B}}{B})^2$ plotted against time after the last ELM for the 55.0 GHz channel ($k_\perp$ ~6-9 cm$^{-1}$).

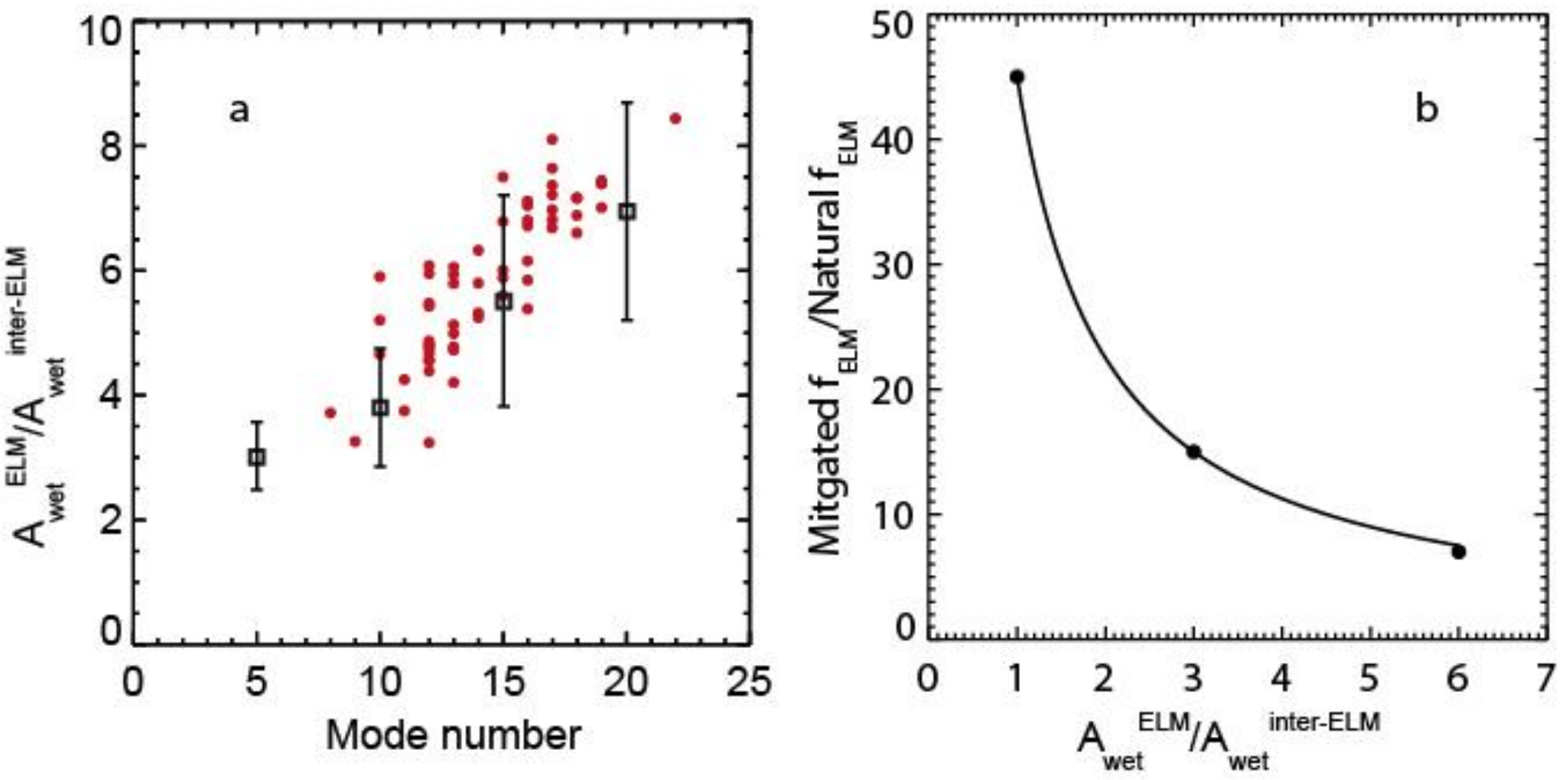

**Figure 8** a) Ratio of ELM to inter-ELM wetted area at the lower outer divertor as a function of the ELM quasi-mode number derived from visible data (red circles). The points with error bars show the modelled ratio. b) Predicted required mitigated ELM frequency as a fraction of the natural ELM frequency for ITER as a function of the increase in wetted area during the ELM compared to inter-ELM.

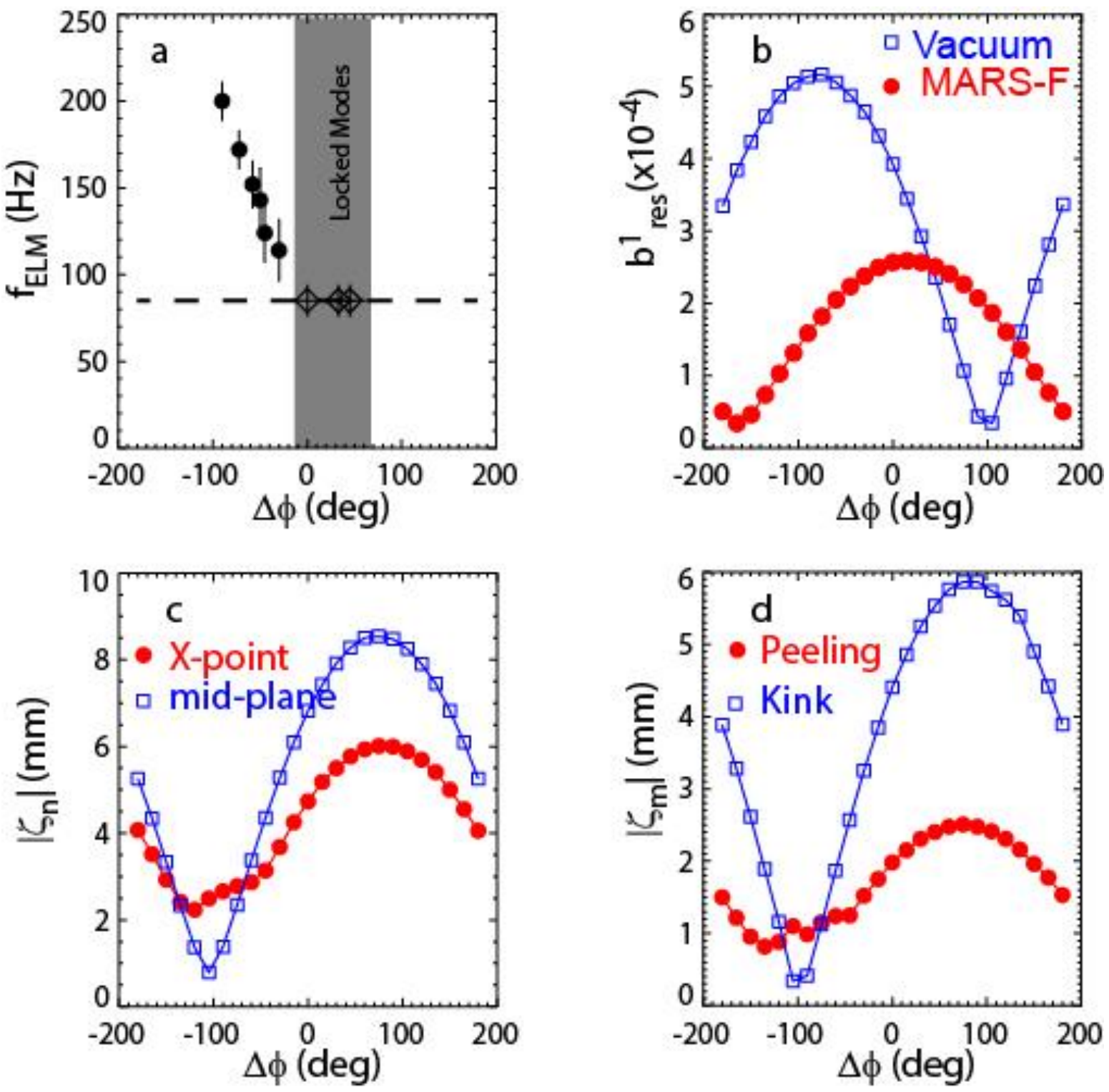

**Figure 9** a) ELM frequency ($f_{ELM}$) b) normalized resonant component of the applied field ($b^r_{res}$) in the vacuum (squares) and including the plasma response (circles) c) the X-point (circles) and mid-plane (squares) displacement and d) the maximum plasma displacement normal to the flux surfaces for the kink (m=1-4) (squares) and peeling (m=6-17) (circles) versus the toroidal phase (Δφ) between the upper and lower row of coils for the RMPs in an n = 3 configuration.

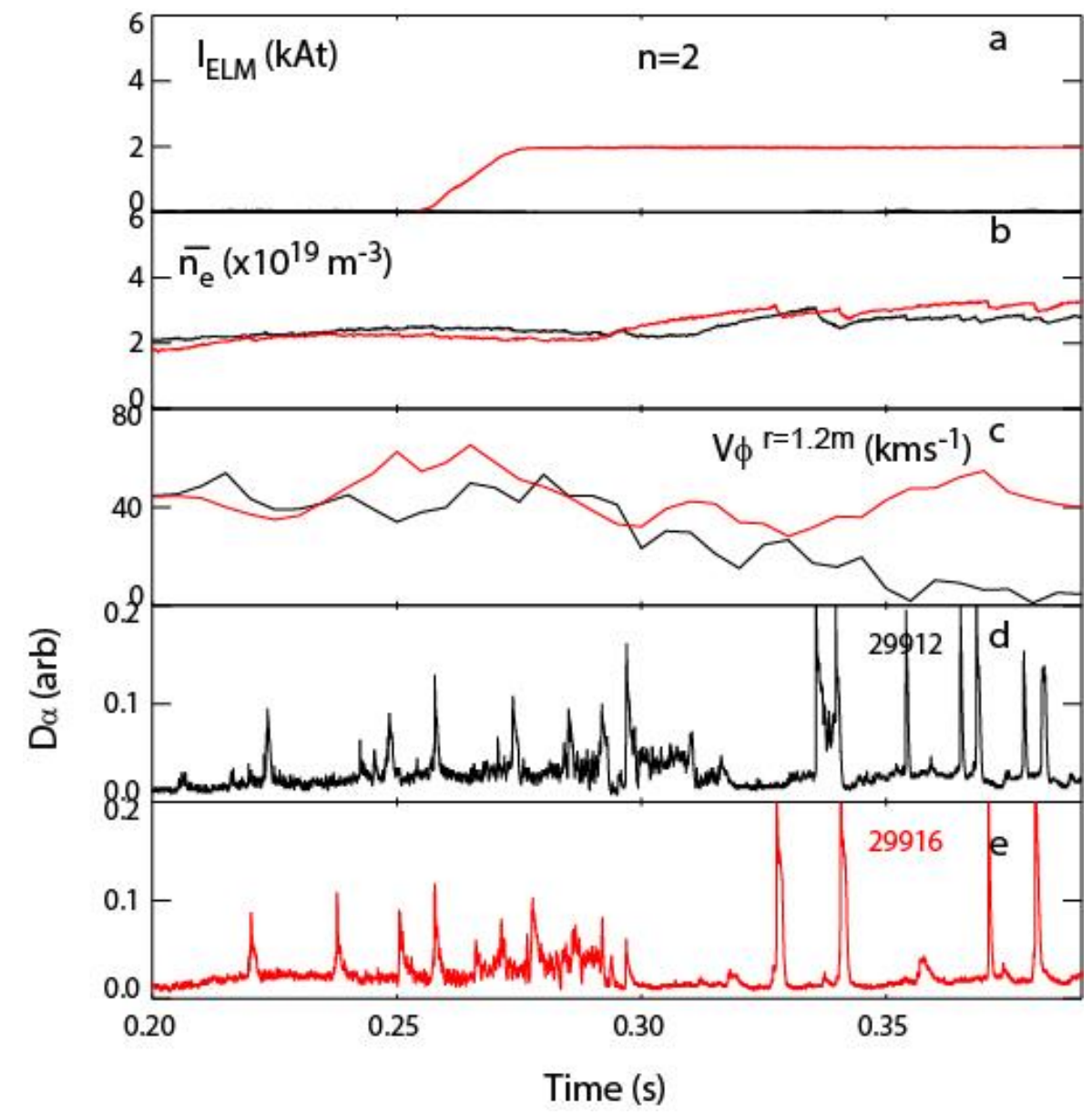

**Figure 10** Time traces of a) the current in the ELM coils ($I_{ELM}$) b) line average density, c) toroidal rotation frequency at r=1.2 m and the divertor $D_\alpha$ light for a shot d) without and e) with n=2 correction.